**AI Agents, Language, Deep Learning and the Next Revolution in Science**

Ke Li[1,2], (like@ihep.ac.cn), Beijiang Liu[1,2] (liubj@ihep.ac.cn), Bruce Mellado[2,3] (bmellado@mail.cern.ch), Changzheng Yuan[1,2] (yuancz@ihep.ac.cn), Zhengde Zhang[1,2] (zdzhang@ihep.ac.cn)

1, Institute of High Energy Physics, Chinese Academy of Sciences, Beijing 100049, China

2, University of Chinese Academy of Sciences, Beijing 100049, China

3, School of Physics and Institute for Collider Particle Physics, University of the Witwatersrand, Johannesburg, Wits 2050, South Africa

**Abstract:** Modern science is reaching a critical inflection point. Instruments across disciplines, from particle physics and astronomy to genomics and climate modeling, now produce data of such scale, diversity, and interdependence that traditional analytical methods can no longer keep pace. This growing imbalance between data generation and data understanding signals the need for a new scientific paradigm. We propose that intelligent, human-supervised AI agents operating over deep-learning algorithms, represent the next evolution of the scientific method. Built upon large language models and multimodal learning, these agents can interpret scientific intent, design and execute analytical workflows, and ensure traceability through domain-specific languages that preserve human oversight and accountability. Particle physics, a historic incubator of computational innovation, offers the ideal testbed for this transition. At the Institute of High Energy Physics of the Chinese Academy of Sciences, the Dr. Sai system embodies this vision, a multi-agent reasoning framework deployed within collider research at the CEPC. This emerging approach does not replace human scientists but extends their cognitive reach, enabling discovery to scale with complexity and redefining how knowledge itself is produced in the age of intelligent machines. The significance of this paradigm transcends particle physics, offering a blueprint for all data-driven sciences facing the same complexity ceiling.

Science is entering an era in which its own success threatens to exceed its methods. Across disciplines, instruments and sensors now produce data so abundant, diverse, and interdependent that traditional analysis can no longer keep pace. The challenge is not merely computational but epistemic: how can understanding scale with complexity? A new paradigm is emerging, one in which intelligent agents assist scientists in reasoning, planning, and executing analyses under human supervision. This evolution, already taking shape in the most data-intensive frontiers such as particle physics, marks a turning point for the scientific method itself.

**The Complexity Ceiling in Data-Intensive Science**

Modern science stands at a crossroads. Across disciplines, from particle physics and astronomy to genomics, neuroscience, and climate modelling, the scale and complexity of data are expanding at an unprecedented pace. Instruments at the frontiers of discovery now generate Petabytes to Exabytes of information annually, capturing phenomena at multiple temporal and spatial scales. These datasets are not only vast in volume but deeply heterogeneous: they integrate structured and unstructured information, span modalities from sensor signals to images and text, and encode intricate correlations that extend across dimensions and domains.

This data explosion has propelled scientific research into new territory, but it has also exposed the limitations of our current methods. The classical paradigm of data analysis, built around manually designed pipelines and domain-specific software frameworks, is increasingly unable to keep pace with the demands of contemporary science. Each new generation of instruments produces data that is larger, more complex, and more interdependent than the one before. Analytical workflows that once sufficed are now brittle, labor-intensive, and difficult to scale. Integrating multimodal datasets, combining, for example, experimental observations with simulation outputs or linking molecular measurements to clinical outcomes remains an arduous task, often requiring bespoke solutions that cannot be easily generalized or reused.

Compounding these technical barriers are mounting human and economic constraints. Sophisticated data analysis today relies on teams of highly specialized experts who must design, implement, and maintain complex software infrastructures. The pool of such expertise is limited and unevenly distributed, and recruitment and retention costs are rising sharply. Even well-resourced institutions struggle to sustain the personnel needed to support large-scale analysis pipelines, while smaller teams and emerging research communities are often excluded altogether. The financial burden of custom programming and analytics, including software development, high-performance computing, and data management, is growing faster than research budgets themselves.

As such, many fields are approaching what might be called a complexity ceiling: a point at which the sheer size, diversity, and dynamism of data, combined with the escalating costs of expertise and infrastructure, threaten to outstrip our collective capacity to interpret it. Unless the way we conduct data analysis itself evolves, the next generation of experiments risks producing insights more slowly, less completely, or not at all.

**Reconfiguring the Scientific Workflow, a Paradigm Shift**

For decades, scientific discovery has followed a linear, human-intensive workflow: researchers formulate hypotheses, design analysis pipelines, code algorithms, and manually validate results. Even as deep learning and automation have accelerated individual stages, the underlying logic has remained unchanged, the scientist directs every computational step. Figure 1. depicts the current paradigm where scientists build and configure multiple independent tools that process data through sequential stages. Each iteration demands significant human effort for configuration and interpretation, emphasizing the inefficiency and fragility of manual analytical pipelines in the era of exponential data growth.

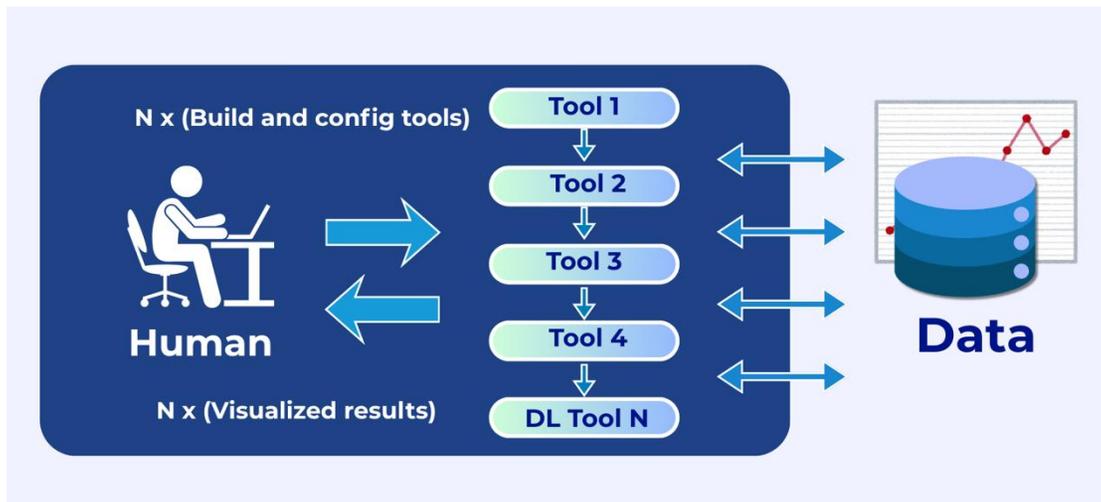

Figure 1. Manual, tool-based analysis where humans configure and manage each stage of the data pipeline, an increasingly unsustainable model as scientific data and complexity expand.

The new paradigm retains this principle but reimagines *how* control is exercised. Rather than performing each operation directly, scientists now guide ensembles of intelligent agents that carry out the technical work under explicit human instruction and oversight.

In this framework, the human remains at the helm, where the AI facilitates cumbersome workflows. The role of the researcher expands from executor to *strategist and curator of knowledge*. Scientists define the objectives, pose the questions, set the constraints, and interpret the outcomes. AI agents, implemented as reasoning systems built on large language models, act as collaborators that understand these objectives, plan analyses, and execute the necessary computational tasks. They handle complexity at scale, yet every action they take originates from, and is ultimately accountable to, human intention.

Central to this shift is precise requirements expression in a human-machine dual-friendly manner. Scientists describe *what* they wish to explore in clear logical statements with low ambiguity. AI agents translate this high-level intent into executable workflows while documenting each decision, ensuring that automation remains traceable, interpretable, and modifiable. A Domain-Specific Language (DSL) for science is a first step, which can be the contract between human understanding and machine execution: a record of purpose, logic, and accountability.

The architecture is multi-agent and modular. Specialized agents address tasks such as data preparation, model selection, uncertainty estimation, and visualization, while reviewer agents verify code and reasoning. Pre-existing, domain-trained deep learning models serve as analytical instruments, sophisticated tools that agents can call upon, but not alter. The human scientist supervises this ecosystem, reviewing outputs, validating reasoning, and refining objectives in an iterative loop of feedback and learning.

Agentic workflows are showing broad promise across the natural sciences: recent surveys document AI agents that generate hypotheses, design and schedule experiments, interface with simulators and lab robots, analyze multimodal data, quantify uncertainty, and iteratively refine protocols in domains spanning life sciences, chemistry, materials science, and physics. Enabled by LLMs, multimodal models, and integrated research platforms, these systems shift AI from point solutions to end-to-end, human-supervised research collaborators. This improves throughput while preserving provenance and interpretability via explicit plans and logs. Early results indicate measurable gains in search efficiency (e.g., molecular and materials discovery), adaptive experimental design, and reproducible analytics pipelines, provided human oversight, safety constraints, and audit trails remain central.

This reconfigured workflow does not displace scientific authority; it redistributes operational load in a sustainable manner. What once demanded large teams managing brittle scripts can now be represented as coherent, reusable workflows aligned with evolving scientific questions. Instead of scaling human labor linearly with data volume, the system scales *cognitively*: agents extend the reach of human reasoning without surrendering control. Discovery thus becomes both faster and more deliberate, an enterprise where AI amplifies curiosity, but the human is in the driver's seat.

Figure 2 illustrates the emerging model of *Human–AI collaboration*. Rather than coding each step, researchers express goals in natural or semi-formal language. AI agents interpret, plan, and execute workflows, while humans remain in control, reviewing visualized results and ensuring interpretability. The DSL acts as the bridge between human intent and machine execution.

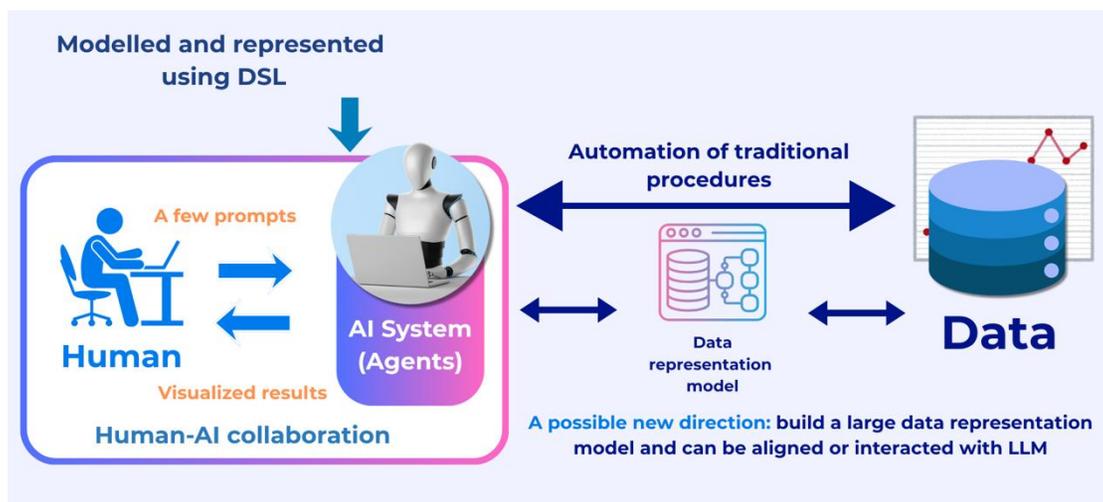

Figure 2. A new paradigm: Human–AI collaboration via domain-specific languages enables automation of traditional analysis while preserving transparency and human oversight.

Crossing the complexity ceiling also redefines *scale*. In the old model, scaling meant distributing computation or personnel; in the new one, it means distributing reasoning itself. Multi-agent systems decompose scientific challenges into coordinated subtasks, synthesize conclusions, and feed structured insights back to their human supervisors. The result is a form of *collective intelligence* that links human judgment, domain-specific models, and machine reasoning within a single transparent framework.

Ultimately, this paradigm restores equilibrium between the growth of data and the growth of understanding. By embedding cognitive collaboration into the research process, scientists can pursue questions that once lay beyond practical reach, tracing uncertainty across billions of events, correlating signals across modalities, or testing hypotheses in real time against streaming observations. The complexity ceiling, once a barrier imposed by the limits of human bandwidth, becomes a moving frontier: one that recedes as new forms of human–AI partnership advances the scope of discovery.

**From Colliders to Cognitive Systems: Physics as the Cradle of Applied Machine Learning**

Few disciplines have confronted complexity as directly, or as early, as particle physics. The search for the Higgs boson, culminating in its discovery at CERN's Large Hadron Collider (LHC) and the 2013 Nobel Prize in Physics, was not only a triumph of experimental science but also a milestone in data-driven discovery. Long before "data science" became a field of its own, collider physicists were already building and deploying the methods that would later define it.

Particle physics was among the first scientific domains outside computer science to embrace machine learning. As early as 1988, physicists began applying neural networks and statistical learning algorithms to particle identification, event classification, and detector calibration. The LEP electron–positron collider, which began operation at CERN in 1989, became the first large-scale scientific experiment to integrate machine learning systematically into its analysis pipelines. By the early 1990s, supervised algorithms were routinely used to separate rare signals from overwhelming backgrounds, a necessity when probing energies where new particles might emerge.

The field's commitment to data-centric methods deepened with the advent of the Tevatron Collider at Fermilab in Chicago and later the Large Hadron Collider at CERN. Each new accelerator generated exponentially larger and more intricate datasets, composed of billions of collision events captured by detectors with millions of readout channels. Extracting meaningful information from this torrent required innovations in data handling, distributed computing, and algorithmic inference. Machine learning became not a tool but an organizing principle of the discipline hardcoded into the DNA of collider physics.

Today, the frontier is shifting once again. The same community that pioneered large-scale data analysis now faces challenges that test the limits of even its own methods. As detector resolutions, event rates, and analysis dimensions expand, traditional machine learning pipelines are strained by scale and complexity. Yet this pressure places collider physics in a uniquely advantageous position. Its decades-long experience at the intersection of computation and discovery make it an ideal testbed for the next leap: the integration of foundation models, large language models, and AI agents into the scientific workflow itself.

In this context, the transition described earlier, from manual pipelines to agent-orchestrated, human-supervised reasoning systems, is not foreign to particle physics; it is a natural evolution. Collider experiments already embody the principles of distributed expertise and collaborative analysis at global scale. Embedding reasoning agents that can interpret goals, plan tasks, and manage analytical complexity represents the next logical step in this lineage. The same impulse that drove the community to invent machine learning for physics in the 1980s now positions it to pioneer the age of cognitive collaboration in science.

As a data-driven Big Science enterprise, accelerator-based particle physics provides a uniquely fertile ecosystem for developing and scaling such an AI-agent framework. Its experimental model already integrates thousands of scientists, layered collaborations, standardized data infrastructures, and transparent governance, precisely the ingredients needed for coordinated human–AI systems. The combination of continuous data streams, complex detector operations, and distributed computational resources offers a natural platform on which reasoning agents can be deployed, tested, and refined in real time. Moreover, the community's long-standing culture of data policies, rigorous validation, and reproducibility aligns perfectly with the requirements of traceable, human-supervised AI. Few other scientific domains possess both the scale and organizational maturity to host an end-to-end, agent-based framework that can evolve from prototype to production within a functioning global experiment.

In this sense, the field that helped launch the era of scientific machine learning is poised once again to lead, this time by pioneering how human and artificial intelligence collaborate in the pursuit of fundamental knowledge. Collider physics not only demonstrates *why* such a paradigm is needed, but *how* it can be realized at *scale*.

**Proof of Principle: Dr. Sai and the New Generation of Colliders**

The next wave of collider experiments, including the Future Circular Collider (FCC-ee) in Europe, the International Linear Collider (ILC) in Japan, and the Circular Electron–Positron Collider (CEPC) in China, heralds not only a leap in experimental precision but also a transformation in how large-scale science is organized and extracts insights from data. These facilities are envisioned to collide electrons and positrons in exceptionally clean conditions, generating datasets of unmatched clarity and diversity. Their environment provides an ideal testbed for implementing the paradigm of human-guided, agent-orchestrated discovery.

Several characteristics make this ecosystem uniquely suited to the transition. The simplicity of $e^+e^-$ interactions, free from the overwhelming hadronic backgrounds of proton colliders, permits the full deployment of deep learning and reasoning-based AI models across every analysis stage. The scale of these projects will mobilize thousands of scientists across continents, forming a natural community to co-develop interoperable digital frameworks, including domain-specific languages, shared datasets, and agent infrastructures. The experiments will produce not only unprecedented volumes but also highly diverse, multimodal data encompassing detector signals, simulation outputs, and theoretical predictions, all of which demand coordinated, adaptive analysis pipelines. Moreover, collider physics has a strong tradition of building ambitious yet cost-effective projects; this culture of disciplined innovation fosters the ideal environment for scalable, transparent AI deployment.

Within this context, the Institute of High Energy Physics (IHEP) of the Chinese Academy of Sciences has developed Dr. Sai, a pioneering proof of principle for the paradigm described above. Dr. Sai is conceived as a *multi-agent scientific intelligence system* that unifies human guidance, language-based reasoning, and domain-specific computation.

Built around a Domain-Specific Language for physics (SaiScript), Dr. Sai enables researchers to describe analytical goals in natural or semi-formal language. AI agents translate these objectives into executable workflows, orchestrating simulation, reconstruction, and statistical inference, while maintaining full traceability of every decision. Each agent has a defined scientific role: one may manage dataset curation, another optimize detector geometry, another monitor systematic uncertainties, and yet another evaluate physical hypotheses. Communication among agents occurs through shared representations of physics logic rather than ad hoc code, ensuring consistency and interpretability.

This architecture transforms the collider environment into a living ecosystem of human–AI collaboration. Scientists remain the ultimate decision-makers: they define the objectives, validate intermediate results, and interpret outcomes. Dr. Sai automates the heavy cognitive and computational lifting, exploring parameter spaces, coordinating workflows, and ensuring reproducibility at scales impossible for manual management.

Figure 3 outlines the operational logic of *Dr. Sai*, a proof-of-principle implementation at the Chinese Academy of Sciences. The system integrates human prompts, knowledge retrieval, and inference layers to generate and execute DSL-based workflows for physics analysis. It represents a prototype for agentic science, where distributed reasoning complements human expertise across data-intensive domains.

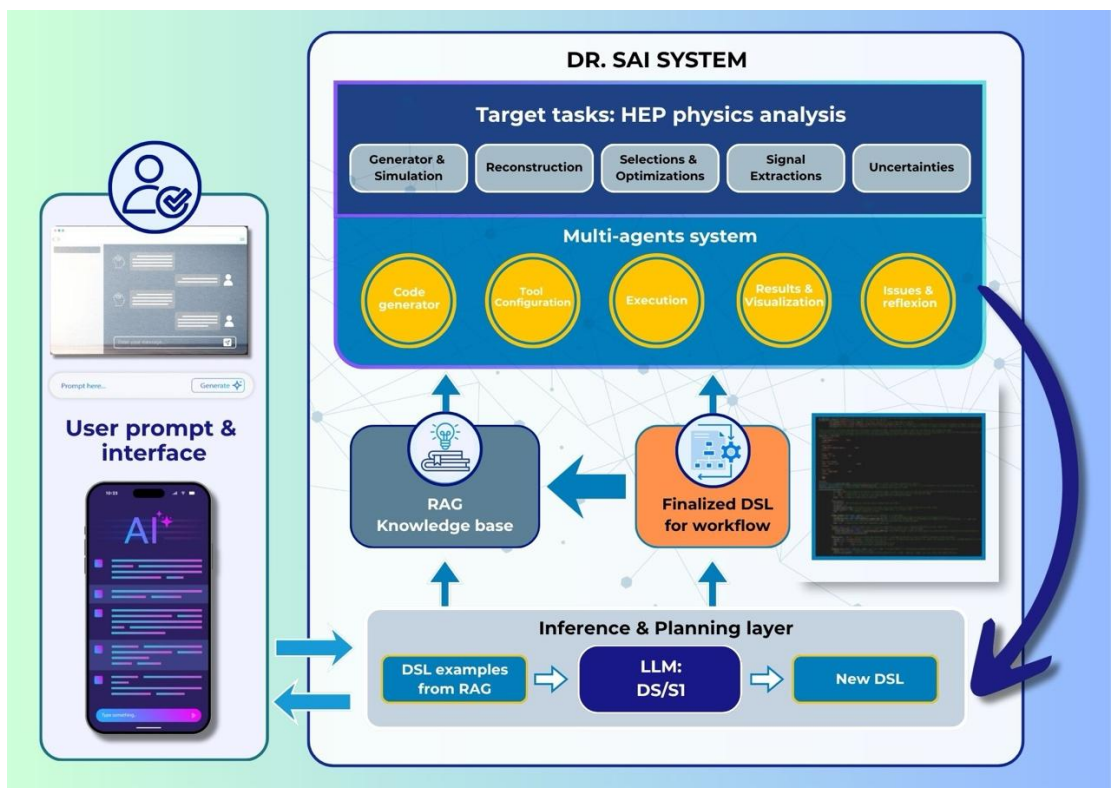

Figure 3. Architecture of Dr. Sai, a modular, multi-agent system integrating reasoning, language, and knowledge retrieval to orchestrate high-energy physics analyses under human supervision.

The deployment of Dr. Sai at the CEPC will mark a decisive milestone, demonstrating that the shift from *data analysis* to *reasoning orchestration* is not aspirational but achievable within a real, high-stakes scientific environment. The field's data intensity, collaborative scale, and culture of precision, now make it the ideal proving ground for agentic science, where deep learning, large language models, and human expertise converge to extend the frontier of discovery. Validation of this framework at scale will show how distributed reasoning can be coordinated across thousands of researchers and Exabytes of data with full transparency and accountability. The resulting architecture will not only accelerate progress in fundamental physics but also provide a transferable blueprint for other data-intensive fields, from climate and materials science to biology and medicine, demonstrating how complex, collaborative research can evolve toward a new, human-supervised, AI-enabled scientific method.